\documentstyle[11pt,newpasp,twoside,epsf,psfig]{article}
\def\edcomment#1{\iffalse\marginpar{\raggedright\sl#1\/}\else\relax\fi}
\reversemarginpar

\begin{document}
\title{Globular Clusters in Early-Type Galaxies with GMOS}
 \author{Terry Bridges, Mike Beasley, Favio Faifer, 
Duncan Forbes, Juan Forte, 
Karl Gebhardt, Dave Hanes, 
Ray Sharples, Steve Zepf} 
\affil{AAO, Swinburne, UNLP/CONICET, Swinburne, 
UNLP/CONICET, \\ U. of Texas, Queen's U.,
Durham, Michigan State U.}


\begin{abstract}

We present recent results from our long-term Gemini/GMOS study of
globular clusters (GCs) in early-type galaxies. 
To date, we have obtained
photometry and spectroscopy for GCs in 
NGCs 3379, 4649, 524, 7332, and IC 1459.
We find a clear bimodality in the NGC 4649 GC colour distribution,
with the fraction of blue/red clusters increasing
with galactocentric radius.  
We derive ages and metallicities for 22 GCs in NGC 3379,
finding that most of the clusters appear old (10$-$15 Gyr);
however, there is a group of
4 metal-rich, younger clusters with ages of 2$-$6 Gyr.
The NGC 3379 GC velocity dispersion decreases with radius, 
as does the
inferred (local) mass-to-light ratio: there is {\it no}
evidence for a dark matter halo in NGC 3379 based on our
GC data.

\end{abstract}

\section{Introduction}

Globular clusters (GCs) are excellent probes of the dynamics,
dark matter content, star-formation histories, and chemical
enrichment of early-type galaxies.  We have embarked upon a 
major programme using the Gemini Multi-Object Spectrograph
(GMOS) on the
8m Gemini telescopes to obtain photometry and spectroscopy of GCs
in $\sim$ 12 early-type galaxies, covering a range of galaxy
type, luminosity, and environment.  In this paper we present
some of the first results from this programme.

\section{Data}

We have been using GMOS on Gemini North (since 2002A) and
South (since 2003B) to obtain spectroscopy for 30$-$100 GCs
in each of NGCs 3379, 4649, 524, 7332, and IC 1459.  GMOS
pre-imaging 
in the g', r', and i' filters for 2$-$3 fields
per galaxy (400$-$800 sec in
each filter) is used to select
GC candidates for follow-up spectroscopy and for photometric
analysis of the GC systems.
The GMOS images
have been reduced using IRAF/DAOPHOT, with photometric 
calibration from HST photometry kindly supplied by Soeren Larsen.
Final GC candidate
lists are determined after rejection of resolved objects, and
objects with colours outside the range of Galactic GCs.  GMOS
multi-slit spectroscopy has been obtained for one or more fields 
in each galaxy, with 25$-$50 slits per field/mask (depending on
the richness of the GC system), and exposure times of 8$-$10 hours
per mask.  Spectroscopic data reduction is done using the 
IRAF/GMOS package.

\section{Globular Cluster Photometry in NGC 4649}

NGC 4649 is a luminous (M$_V$=$-$22.4) Virgo elliptical.  We
present preliminary photometric results for $\sim$ 2000 unresolved
objects with i' $<$ 25 and 0.5 $<$ (g'-i') $<$ 1.5 (see 
Forbes et al. (2004) for a fuller account, and
Bridges et al. (2004) for NGC 4649 GC spectroscopy). 
The NGC 4649 GC
colour distribution is clearly bimodal, with peaks at 
(g'-i') = $\sim$ 0.9 and 1.15.  The fraction of
blue/red clusters increases with radius, as found in
other early-type galaxies.  

\section{Globular Cluster Spectroscopy in NGC 3379}

NGC 3379 is a less luminous elliptical in the nearby Leo group.
We have spectra for 22 NGC 3379 GCs in a 10 hour
GMOS exposure with 5 \AA\ resolution and coverage
from $\sim$ 4000$-$7000 \AA\ .  {\bf Figure 1} shows the
ages and metallicities of 18 GCs, based on the comparison of
line-strengths with stellar population synthesis models.  Most 
of the GCs are old, with ages $>$ 10 Gyr, but there is a group
of $\sim$ 4 metal-rich, younger GCs with ages of 2$-$6 Gyr.  
{\bf Figure 2} shows that both the GC velocity dispersion
(top panel) and the local M/L (bottom panel) decrease outwards.
A preliminary analysis (see Beasley et al. 2004) 
shows that {\it there is
no evidence for a dark-matter halo in NGC 3379}, a very interesting
result also seen in the PNe kinematics (Romanowsky et al.
2003, Science, 301, 1696).

\noindent
\begin{minipage}{2.0in}
\psfig{file=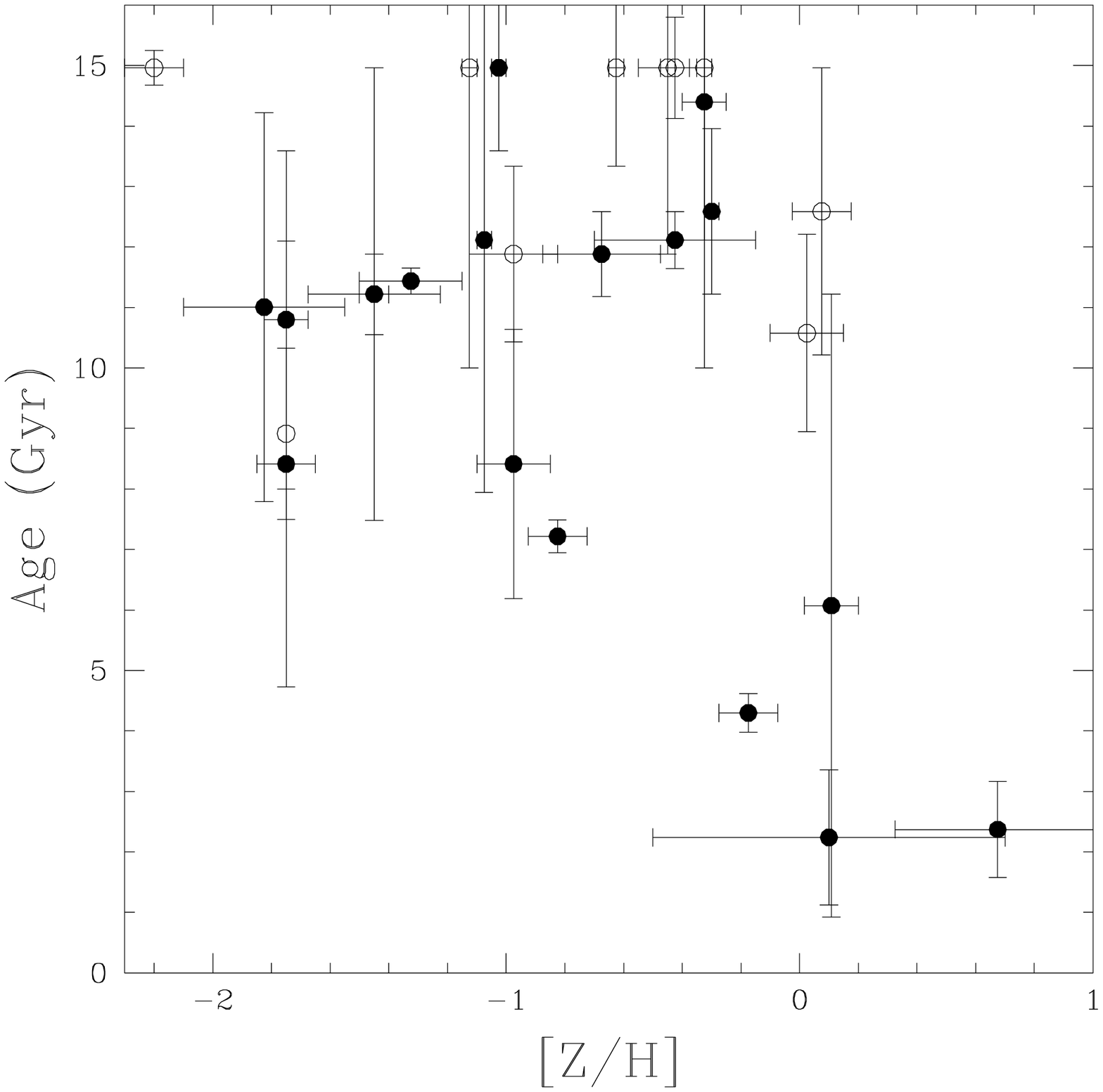,angle=0,width=2.5in,height=2.5in}
\end{minipage}
\hspace*{0.3in}
\begin{minipage}{2.0in}
\psfig{file=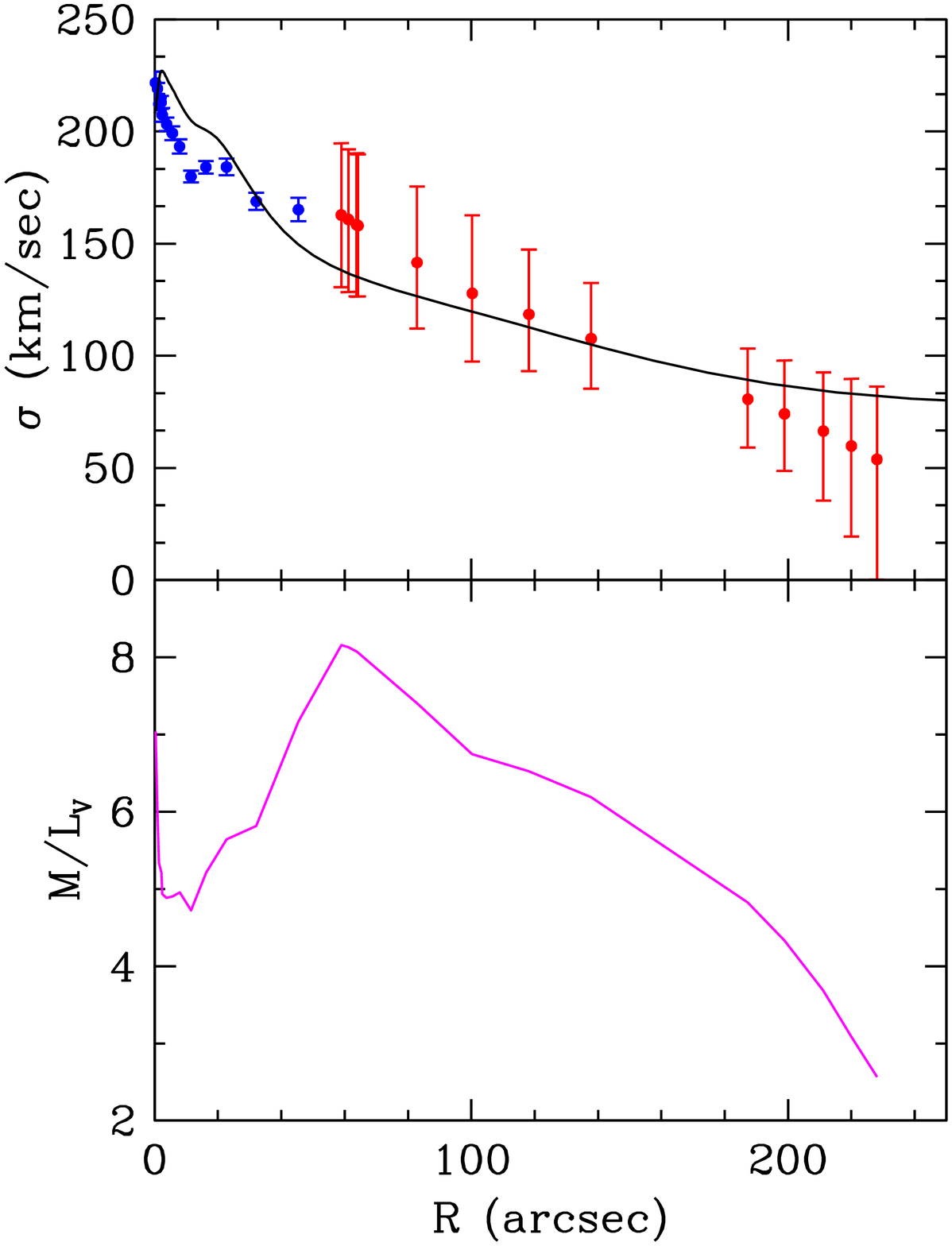,angle=0,width=2.5in,height=2.5in}
\end{minipage}

\bigskip

{\bf Figure 1 (left):} Ages/metallicities for NGC 3379
GCs (filled circles) and MW GCs (open circles), from
our GMOS spectra. 
{\bf Figure 2 (right):}
{\it Top:} velocity dispersion of NGC 3379 stars (small errors)
and GCs (larger errors); the solid line is a constant M/L fit.
{\it Bottom:} The local M/L for NGC 3379.

\bigskip

\acknowledgements

We are very grateful to Soeren Larsen for all of his help with our photometric
calibration, and to the Gemini staff, particularly Inger Jorgensen, for
obtaining such nice data for us.

\end{document}